\documentclass[aps,prl,reprint,showpacs,superscriptaddress]{revtex4-1}

\usepackage{amssymb}
\usepackage{amsmath}
\usepackage{graphicx}

\begin{document}


\title{Ultrafast tuneable optical delay line based on indirect photonic
transitions}


\author{Daryl M. Beggs}
\email[]{beggs@amolf.nl}
\thanks{These authors contributed equally.}
\affiliation{FOM Institute AMOLF, Science Park 104, 1098XG Amsterdam,
The Netherlands}

\author{Isabella H. Rey}
\thanks{These authors contributed equally.}
\affiliation{SUPA, School of Physics and Astronomy, University of St Andrews,
St Andrews, Fife KY16 9SS, UK}

\author{Tobias Kampfrath}
\affiliation{FOM Institute AMOLF, Science Park 104, 1098XG Amsterdam,
The Netherlands}
\affiliation{Fritz Haber Institute of the Max Planck Society, Faradayweg
4-6,
14195 Berlin, Germany}

\author{Nir Rotenberg}
\affiliation{FOM Institute AMOLF, Science Park 104, 1098XG Amsterdam,
The Netherlands}

\author{L. Kuipers}
\affiliation{FOM Institute AMOLF, Science Park 104, 1098XG Amsterdam,
The Netherlands}

\author{Thomas F. Krauss}
\affiliation{SUPA, School of Physics and Astronomy, University of St Andrews,
St Andrews, Fife KY16 9SS, UK}

\date{\today}

\begin{abstract}
We introduce the concept of an indirect photonic transition and demonstrate
its use in a dynamic delay line to alter the group velocity of an optical
pulse. Operating on an ultrafast time scale, we show continuously tuneable
delays of up to 20~ps, using a slow light photonic crystal waveguide only
300~$\mu$m in length. Our approach is flexible, in that individual pulses
in a pulse stream can be controlled independently, which we demonstrate by
operating on pulses separated by just 30~ps. The two-step indirect
transition is demonstrated here with a 30\% conversion efficiency.
\end{abstract}


\maketitle



Indirect transitions are well-known phenomena in solid state physics, most
notably for electrons inside a semiconductor material, whereby the
simultaneous absorption of a photon and a phonon results in a change of
both energy and momentum \cite{Ashcroft_SolidState1976}. The corresponding
optical analogue, the indirect \textit{photonic} transition, has so-far
been elusive. Several schemes have been proposed
\cite{Winn_Ippen_PRB59:1551_1999, Yu_Fan_NP3:91_2009}, that are based on
the simultaneous temporal and spatial modulation of the refractive index,
which induces the transition between two photonic states of different
frequency and wavevector. Such an indirect photonic transition has, to our
knowledge, not yet been observed, and the closest related experimental
demonstrations are those of direct interband photonic transitions 
\cite{Dong_Lipson_PRL100:033904_2008}.

\begin{figure}
\includegraphics[width=8.6cm]{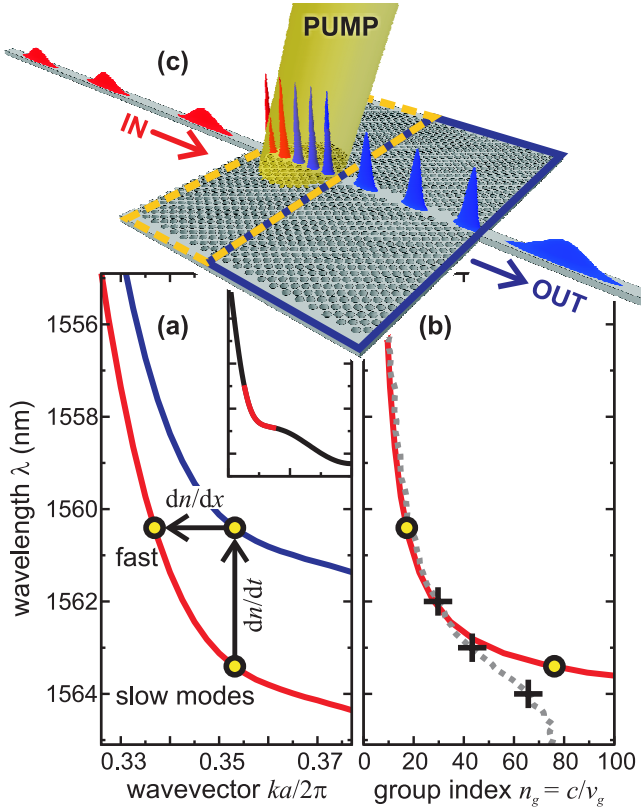}
\caption{\label{fig1}(color online) Schematic of photonic indirect
transition in a photonic crystal waveguide.
(a) Calculated dispersion of waveguide mode in ground state (red) and
excited state (blue), with $\Delta n = -7\times10^{-3}$. Yellow
dots and arrows indicate the two-step transition of a pulse from slow to
fast-light. Inset: ground state dispersion calculated over a wider
wavelength range, with area of detail in main figure indicated in red.
(b) Calculated (red) and measured (dashed grey) group index spectra for the
photonic crystal waveguide. Yellow dots indicate start/end points of
indirect transition in (a). Black crosses indicate wavelengths 
used in the experiments.
(c) Illustration of the delay line. A
red pulse is input from the left, where it undergoes adiabatic wavelength
conversion in the tuning region (outlined in yellow) that is excited by the
ultrafast pump-pulse. The blue pulse then travels through the static delay
region (outlined in blue) at a speed which depends on its final
wavelength.}
\end{figure}

In this Letter, we propose and 
demonstrate a simplified
version of an intraband indirect photonic transition, whereby the frequency
and the wavevector are altered in a two-step process. We also show that
such an approach can form the core of an ultrafast, continuously tuneable
delay line capable of bit-by-bit control, which we demonstrate by
independently manipulating one pulse from a chain of two that are separated
by just 30~ps.

Our structure is designed to implement the transition of a light pulse from
a slow state into a fast state of a photonic crystal waveguide, as
illustrated in the dispersion diagram of Fig.\ \ref{fig1}(a) and the group
index curve $n_g=c/v_g$ of Fig.\ \ref{fig1}(b). Such a transition requires
a change in both frequency and wavevector, which can be provided by a
stimulus that involves temporal and spatial perturbations of the waveguide,
respectively. We realized this scheme
by borrowing concepts from adiabatic light control
\cite{Yanik_Fan_PRL92:083901_2004, Beggs_Kampfrath_PRL108:033902_2012, 
Kampfrath_Kuipers_PRA81:043837_2010,  Upham_Noda_APEX3:062001_2010}
and fibre optics \cite{Nuccio_Willner_OL35:1819_2010,
Sharping_Gaeta_OE13:7872_2005}.

As illustrated 
in Fig.\ \ref{fig1}(c), our fabricated structure is
a single 300~$\mu$m long modified version of a W1 photonic crystal
waveguide \cite{Li_Krauss_OE16:6227_2008}, of period $a=416$~nm and hole
diameter 249~nm, designed to exhibit the dispersion curve indicated by the
red line of Fig.\ \ref{fig1}(a). Typically, such modifications aim to
eliminate group velocity dispersion, but here we introduce it in a
controlled fashion. The first 80~$\mu$m is used as a tuning section (yellow
box of Fig.\ \ref{fig1}(c)), where an input signal pulse undergoes
adiabatic frequency conversion. This process is triggered by a 
pump-pulse that decreases the refractive index of the silicon through the 
ultrafast generation of free carriers \cite{Soref_Bennett_IEEEJQE23:123_1987} 
with  lifetimes \raise.17ex\hbox{$\scriptstyle\mathtt{\sim}$}200~ps, which
blue-shift the dispersion curve (blue curve in Fig.\ \ref{fig1}(a)). The
signal pulse is also blue-shifted
\cite{Kampfrath_Kuipers_PRA81:043837_2010}, as the photon momentum 
$\hbar k= \hbar n \omega/c$ must be conserved: it is the generation of 
free carriers 
 which provides the dynamics $\mathrm{d}n/\mathrm{d}t$ necessary for a
vertical shift on the dispersion diagram. The interface between the 
pumped and unpumed region (blue box of Fig.\ \ref{fig1}(c)) 
provides the spatial modulation $\mathrm{d}n/\mathrm{d}x$ necessary
for the horizontal transition, altering the signal wavevector. 
The remainder of the 
 untuned section is there to ensure that the speed difference between the 
 pumped and unpumped pulses can manifest itself as a change in delay.

\begin{figure}
\includegraphics[width=8.6cm]{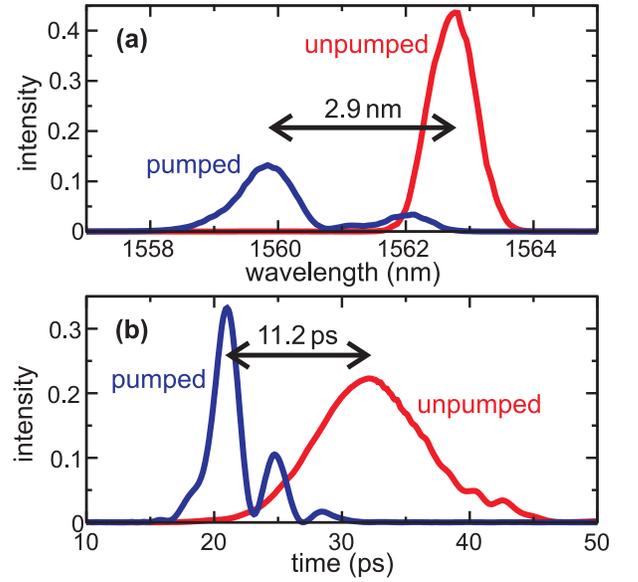}
\caption{\label{fig2}(color online) Wavelength conversion and delay of
pulses. (a) Unpumped (red) and pumped (blue) pulses in the wavelength
domain. The pumped pulse is blue-shifted by 2.9~nm w.r.t.\ the unpumped
one.
(b) The same unpumped (red) and pumped (blue) pulses from (a), but in the
time domain. The wavelength converted pumped pulse arrives 11.2~ps 
(or 2.5 pulse widths) earlier than the unpumped pulse.}
\end{figure}

The occurrence of an indirect transition in our 
waveguide can be verified by recording the wavelength and the arrival time of 
the output signal pulse with and without pumping. The pump pulses are 
generated in a Ti:sapphire laser (centre wavelength 810~nm, duration 100~fs, 
repetition rate 80~MHz) and focussed onto the surface of the sample. A slit 
is used in the far-field to ensure the exclusive and homogeneous excitation 
of the waveguide in the 
tuning region. 
Synchronised probe pulses (1550~nm, 180~fs) 
are obtained from the same laser system: after passing through an 
optical parametric oscillator, they are coupled into the waveguide, 
with a mechanical delay stage used to set the timing between the pump 
and probe pulses. By measuring the waveguide 
time-dependent transfer function using the broadband 180-fs probe pulse, we
have determined the response to an input pulse of duration 4.7~ps and
centre wavelength 1563~nm. This approach
\cite{Lepetit_Joffre_JOSAB12:2467_1995, Kampfrath_Kuipers_OL34:3418_2009}
simply assumes that the waveguide is linear from the weak probe point of
view, which we have verified experimentally. The extracted spectral output
of this pulse from a ground state waveguide is shown by the red curve in
Fig.\ \ref{fig2}(a). The transmission is measured as -3~dB (50\%), with the 
primary losses due to propagation in the slow light regime
(\raise.17ex\hbox{$\scriptstyle\mathtt{\sim}$}10~dB/mm). When tuned, the
pumped input pulse is adiabatically shifted by 2.9~nm 
to a new centre wavelength of 1559.9~nm (blue curve in Fig.\ \ref{fig2}(a)) 
with a conversion efficiency of -5~dB (30\%), manifesting the vertical 
transition in the band diagram. The losses in the conversion process arise 
from free-carrier absorption. The dispersion at the original wavelength 
results in a small fraction of the probe not being fully contained in the 
tuning region during pumping, thus leaving some residual light around 1562~nm.

In the time-domain, the frequency conversion translates into a change of
the delay accumulated in the 220~$\mu$m long untuned section (Fig.\
\ref{fig2}(b)).  
The unpumped pulse (red)
travels 
with a group velocity of $c/39$, and exits
after 32.3~ps. The blue-shifted pumped pulse, however, travels with a
higher group velocity of $c/17$, exiting the waveguide 11.2~ps (or 
2.5 pulse-widths) earlier than its unpumped counterpart, with -1.5~dB
propagation loss (\raise.17ex\hbox{$\scriptstyle\mathtt{\sim}$}5~dB/mm) in
the fast light regime. This difference in delay results from the change of
the optical mode of the signal, and together with the measured wavelength
conversion confirms the occurrence of an indirect transition. The 
group velocity dispersion at the original frequency accounts for the
increased width of the unpumped pulse in the time-domain.

The physical mechanism of our two-step indirect transition should not be
directly compared to that of an electron in a semiconductor, which rather
finds its counterpart in the one-step process described in refs.\
\cite{Winn_Ippen_PRB59:1551_1999, Yu_Fan_NP3:91_2009}. Our case is more
similar to that of an electron experiencing a dynamic change in its atomic
potential. As the changes in our waveguide are adiabatic
\cite{Kampfrath_Kuipers_PRA81:043837_2010}, the signal pulse is
continuously ``guided'' by the changes in the photonic band structure. The
final result is an indirect transition, or adiabatic conversion, from an
initial slow mode to a final fast mode of the same photonic band.

We adopt this approach, rather than the one-step photonic transition 
of ref.\ \cite{Yu_Fan_NP3:91_2009} 
for two reasons. First, we are
able to use a homogeneous pump profile to excite the carriers, while a
one-step transition requires a more complicated spatial profile that breaks
the symmetry of the waveguide. Second, our approach is not limited by the
modulation speed. We are limited instead by the magnitude of the
carrier-induced 
index change, which allows us to achieve
frequency shifts on the order of 400 GHz 
\cite{Kampfrath_Kuipers_PRA81:043837_2010}, rather than 20 GHz as was
theoretically shown in ref.\ \cite{Yu_Fan_NP3:91_2009}.

Our photonic crystal 
therefore acts as a tuneable delay line. 
Whereas previous quasi-static
examples \cite{Vlasov_McNab_N438:65_2005,
Melloni_Martinelli_OL33:2389_2008, Adachi_Baba_IEEEJSTQE16:192_2010} 
used the thermo-optic effect to tune the 
\textit{waveguide} from the fast  
to the slow light regime, we use
ultrafast generation of carriers to tune the \textit{pulse}, converting it
from slow to fast light, and offering the possibility of true bit-by-bit
dynamic control of individual pulses with high efficiency and flexibility.
The tuning section 
 only needs to be as long as the pulse
itself, whereas the delay function is performed by the following static
section of arbitrary length. In contrast, conventional tuneable delay lines
necessitate the tuning of the entire delay section, which incurs higher
energy requirements in the case of thermo-optic tuning, or higher losses in
the case of free-carrier tuning.

Note that the indirect transition as described above helps to favourably
balance the two main loss mechanisms in the delay line. The pumped pulses
are absorbed when they overlap with the free-carriers in the tuning
section, but travel faster in the delay section, with lower propagation
losses \cite{OFaolain_Krauss_OE18:27627_2010}. In contrast, the unpumped
pulses travel in the lossier slow light regime, but have no conversion
losses. Dispersion compensation using suitably engineered waveguide
sections can be used to eliminate the broadening of slow pulses, which is
not an intrinsic limitation of our method.

\begin{figure}
\includegraphics[width=8.6cm]{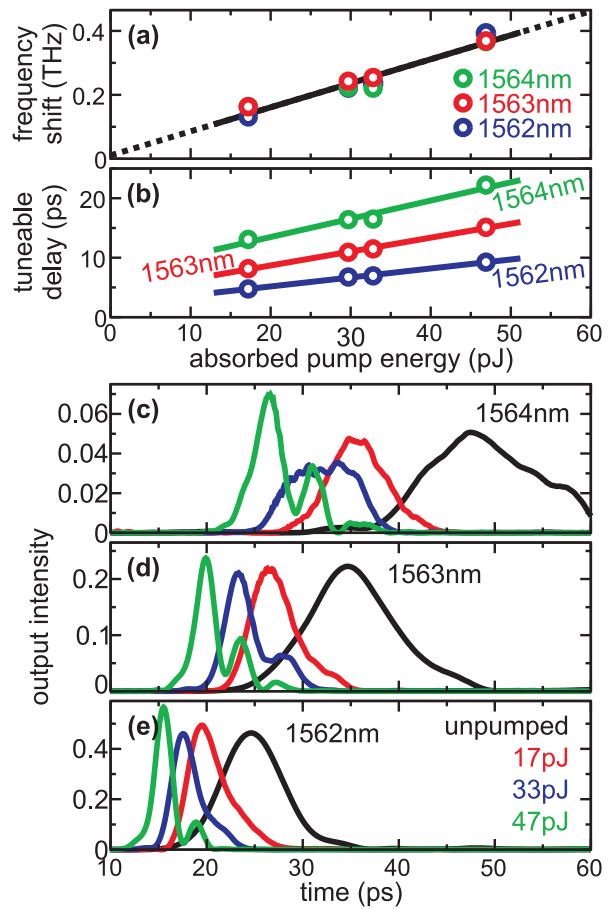}
\caption{\label{fig3}(color online) Continuously tuneable delays via pump
power and wavelength dependence.
(a) Frequency shift as a function of absorbed pump energy for pump pulses
with starting wavelengths 1562~nm (blue), 1563~nm (red) and 1564~nm
(green). The frequency shift is directly proportional to the absorbed pump
energy, as shown by the linear fit (black lines). (b) The measured delays
for the same pulses as in (a). (c-e) The time-domain output of the pulses
from (a) and (b) for starting wavelengths 1564~nm (c), 1563~nm (d) and
1562~nm (e). The colours indicate the absorbed pump energies.}
\end{figure}

The delays demonstrated in Fig.\ \ref{fig2} are continuously tuneable,
either via the power of the pump, 
or by changing the input wavelength,
as shown in Fig.\ \ref{fig3}. The frequency shift is a linear function of
the pump energy (Fig.\ \ref{fig3}(a)) and independent of starting
wavelength. This can be readily understood: the number of free-carriers
generated is directly proportional to the absorbed energy, and the
refractive index change is proportional to the carrier density
\cite{Soref_Bennett_IEEEJQE23:123_1987}. The tuneable delay (Fig.\
\ref{fig3}(b)) depends on the local slope of the group velocity spectrum
(grey line in Fig.\ \ref{fig1}(b)) at the starting wavelength, 
and the corresponding output pulses are
shown in Fig.\ \ref{fig3}(c-e). For example, for a wavelength of 1564~nm,
an absorbed pump-energy of 47~pJ gives a wavelength shift of 3~nm
(0.37~THz) and a delay of 22.1~ps ($>4$ pulse widths).

\begin{figure}
\includegraphics[width=8.6cm]{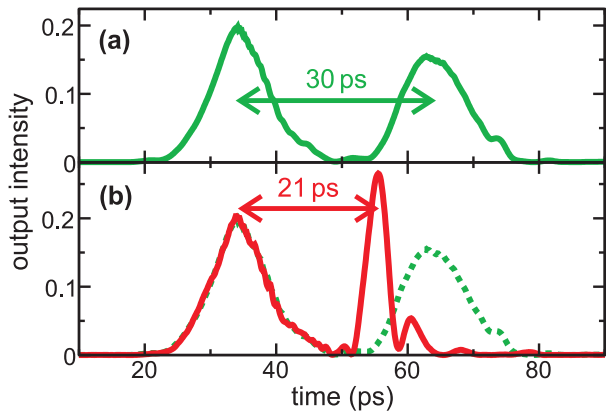}
\caption{\label{fig4}(color online) Tuneable single-bit manipulation in a
train of two pulses. Measured time domain output for two pulses input 30~ps
apart. (a) Neither pulse is pumped, and the two pulses are output still
separated by 30~ps. (b) The second pulse only is pumped (red), and it is
manipulated independently of the first: it is output 9~ps earlier than its
unpumped counterpart, resulting in a new separation of 21~ps. Unpumped
output from (a) is shown (dotted green line) for comparison.}
\end{figure}

Finally, in order to demonstrate the potential of our ultrafast approach
for bit-by-bit control of individual pulses, we have used two pulses of
equal amplitude and centre wavelength of 1563~nm, duration 4.7~ps and
separation 30~ps. Figure \ref{fig4}(a) shows the output when both pulses
pass through the waveguide unpumped: the 
pulses travel in the ground state
and exit the waveguide still separated by 30~ps. Overlapping the second
probe-pulse with the pump in the tuning region allows for it to be
manipulated independently of the first. The first pulse 
exits in the ground state,
whereas the second pulse undergoes frequency conversion and delay. Figure
\ref{fig4}(b) shows that the delay of the first pulse is not altered,
whereas the second pulse exits 9~ps earlier than its unpumped counterpart, 
and the two pulses emerge just 21~ps apart.

We briefly note that optical buffers based on high-Q cavities have also
been demonstrated \cite{Xu_Lipson_NP3:406_2007, 
Tanabe_Kuramochi_PRL102:043907_2009} 
using ultrafast tuning of the cavity resonance or Q-factor to trap and release
pulses. However, such schemes suffer from limited bandwidth (the storage
times are limited by the photon lifetime, necessitating high-Q cavities)
and are much less flexible, as they only allow the delay of a single bit at
a time. The loading of a pulse blocks the device until its release, such
that the delay performance represents the intrinsic limitation on the
dead-time. Our 
indirect transition device, in contrast, is
scalable, in that the 
amount of delay available
is simply related to the length of the waveguide (so the ultimate
limitation is given by the propagation loss). The dead-time is related to
the system recovery after carrier injection, which in principle can be
reduced to tens of ps, as discussed below.

In conclusion, we have demonstrated the occurrence of a two-step indirect
photonic transition based on adiabatic tuning in a photonic crystal
waveguide, which we have applied in the realisation of an ultrafast,
continuously tuneable integrated delay line. The indirect transition is
achieved with a wavelength conversion and dispersion approach and has the
major advantage of scaleability, as only the waveguide section containing
the pulse is pumped. We have demonstrated a tuneable delay of 20~ps,
corresponding to \raise.17ex\hbox{$\scriptstyle\mathtt{\sim}$}4 pulse
lengths, but suitably dispersion engineered and longer waveguides could
achieve much longer fractional delays for the same pump power. The
ultrafast nature of the tuneable delay is demonstrated by manipulating
individual pulses of a pulse-stream, and our geometry 
could be used to manipulate multiple pulses at once. The major limitation
of the current device is the dead-time,  
given by the carrier lifetime 
of
\raise.17ex\hbox{$\scriptstyle\mathtt{\sim}$}200~ps, which could be reduced
by an order of magnitude by sweeping out the carriers via a reverse-biased
pn junction. The scheme is fully compatible with silicon optoelectronics,
as the required wavelength shift could also be produced by carrier
injection/depletion \cite{Nguyen_Baba_OE19:13000_2011,
Tanabe_Notomi_OL35:3895_2010}, removing the need for an external pump
laser.

Hence, our indirect-transition based integrated tuneable delay line
provides an important missing piece of the silicon photonics puzzle,
bringing the full implementation of all-optical networks on silicon chips a
step closer towards reality \footnote{During the reviewing process, 
we became aware of arXiv:1110.5337v1, where an alternative method 
for achieving an indirect photonic transition \cite{Yu_Fan_NP3:91_2009} 
has been demonstrated.}.

\begin{acknowledgments}
The authors 
acknowledge the support of EPSRC through the UK Silicon
Photonics project, and the support of the Smart Mix Programme of the
Netherlands Ministry of Economic Affairs and the Netherlands Ministry of
Education, Culture and Science. This work is also supported by NanoNextNL 
of the Government of the Netherlands and 130 partners, and part of the 
research program of the Stichting voor Fundamenteel Onderzoek der Materie 
(FOM), which is financially supported by the Nederlandse organisatie voor
Wetenschappelijk Onderzoek (NWO).
\end{acknowledgments}


\begin{thebibliography}{23}%
\makeatletter
\providecommand \@ifxundefined [1]{%
 \@ifx{#1\undefined}
}%
\providecommand \@ifnum [1]{%
 \ifnum #1\expandafter \@firstoftwo
 \else \expandafter \@secondoftwo
 \fi
}%
\providecommand \@ifx [1]{%
 \ifx #1\expandafter \@firstoftwo
 \else \expandafter \@secondoftwo
 \fi
}%
\providecommand \natexlab [1]{#1}%
\providecommand \enquote  [1]{``#1''}%
\providecommand \bibnamefont  [1]{#1}%
\providecommand \bibfnamefont [1]{#1}%
\providecommand \citenamefont [1]{#1}%
\providecommand \href@noop [0]{\@secondoftwo}%
\providecommand \href [0]{\begingroup \@sanitize@url \@href}%
\providecommand \@href[1]{\@@startlink{#1}\@@href}%
\providecommand \@@href[1]{\endgroup#1\@@endlink}%
\providecommand \@sanitize@url [0]{\catcode `\\12\catcode `\$12\catcode
  `\&12\catcode `\#12\catcode `\^12\catcode `\_12\catcode `\%12\relax}%
\providecommand \@@startlink[1]{}%
\providecommand \@@endlink[0]{}%
\providecommand \url  [0]{\begingroup\@sanitize@url \@url }%
\providecommand \@url [1]{\endgroup\@href {#1}{\urlprefix }}%
\providecommand \urlprefix  [0]{URL }%
\providecommand \Eprint [0]{\href }%
\providecommand \doibase [0]{http://dx.doi.org/}%
\providecommand \selectlanguage [0]{\@gobble}%
\providecommand \bibinfo  [0]{\@secondoftwo}%
\providecommand \bibfield  [0]{\@secondoftwo}%
\providecommand \translation [1]{[#1]}%
\providecommand \BibitemOpen [0]{}%
\providecommand \bibitemStop [0]{}%
\providecommand \bibitemNoStop [0]{.\EOS\space}%
\providecommand \EOS [0]{\spacefactor3000\relax}%
\providecommand \BibitemShut  [1]{\csname bibitem#1\endcsname}%
\let\auto@bib@innerbib\@empty
\bibitem [{\citenamefont {Ashcroft}\ and\ \citenamefont
  {Mermin}(1976)}]{Ashcroft_SolidState1976}%
  \BibitemOpen
  \bibfield  {author} {\bibinfo {author} {\bibfnamefont {N.~W.}\ \bibnamefont
  {Ashcroft}}\ and\ \bibinfo {author} {\bibfnamefont {N.~D.}\ \bibnamefont
  {Mermin}},\ }\href@noop {} {\emph {\bibinfo {title} {Solid State Physics}}}\
  (\bibinfo  {publisher} {Holt-Saunders},\ \bibinfo {address} {Philadelphia},\
  \bibinfo {year} {1976})\BibitemShut {NoStop}%
\bibitem [{\citenamefont {Winn}\ \emph {et~al.}(1999)\citenamefont {Winn},
  \citenamefont {Fan}, \citenamefont {Joannopoulos},\ and\ \citenamefont
  {Ippen}}]{Winn_Ippen_PRB59:1551_1999}%
  \BibitemOpen
  \bibfield  {author} {\bibinfo {author} {\bibfnamefont {J.~N.}\ \bibnamefont
  {Winn}}, \bibinfo {author} {\bibfnamefont {S.}~\bibnamefont {Fan}}, \bibinfo
  {author} {\bibfnamefont {J.~D.}\ \bibnamefont {Joannopoulos}}, \ and\
  \bibinfo {author} {\bibfnamefont {E.~P.}\ \bibnamefont {Ippen}},\ }\href
  {\doibase 10.1103/PhysRevB.59.1551} {\bibfield  {journal} {\bibinfo
  {journal} {Phys. Rev. B}\ }\textbf {\bibinfo {volume} {59}},\ \bibinfo
  {pages} {1551} (\bibinfo {year} {1999})}\BibitemShut {NoStop}%
\bibitem [{\citenamefont {Yu}\ and\ \citenamefont
  {Fan}(2009)}]{Yu_Fan_NP3:91_2009}%
  \BibitemOpen
  \bibfield  {author} {\bibinfo {author} {\bibfnamefont {Z.}~\bibnamefont
  {Yu}}\ and\ \bibinfo {author} {\bibfnamefont {S.}~\bibnamefont {Fan}},\
  }\href {\doibase 10.1038/nphoton.2008.273} {\bibfield  {journal} {\bibinfo
  {journal} {Nature Photon.}\ }\textbf {\bibinfo {volume} {3}},\ \bibinfo
  {pages} {91} (\bibinfo {year} {2009})}\BibitemShut {NoStop}%
\bibitem [{\citenamefont {Dong}\ \emph {et~al.}(2008)\citenamefont {Dong},
  \citenamefont {Preble}, \citenamefont {Robinson}, \citenamefont
  {Manipatruni},\ and\ \citenamefont
  {Lipson}}]{Dong_Lipson_PRL100:033904_2008}%
  \BibitemOpen
  \bibfield  {author} {\bibinfo {author} {\bibfnamefont {P.}~\bibnamefont
  {Dong}}, \bibinfo {author} {\bibfnamefont {S.~F.}\ \bibnamefont {Preble}},
  \bibinfo {author} {\bibfnamefont {J.~T.}\ \bibnamefont {Robinson}}, \bibinfo
  {author} {\bibfnamefont {S.}~\bibnamefont {Manipatruni}}, \ and\ \bibinfo
  {author} {\bibfnamefont {M.}~\bibnamefont {Lipson}},\ }\href {\doibase
  10.1103/PhysRevLett.100.033904} {\bibfield  {journal} {\bibinfo  {journal}
  {Phys. Rev. Lett.}\ }\textbf {\bibinfo {volume} {100}},\ \bibinfo {pages}
  {033904} (\bibinfo {year} {2008})}\BibitemShut {NoStop}%
\bibitem [{\citenamefont {Yanik}\ and\ \citenamefont
  {Fan}(2004)}]{Yanik_Fan_PRL92:083901_2004}%
  \BibitemOpen
  \bibfield  {author} {\bibinfo {author} {\bibfnamefont {M.~F.}\ \bibnamefont
  {Yanik}}\ and\ \bibinfo {author} {\bibfnamefont {S.}~\bibnamefont {Fan}},\
  }\href {\doibase 10.1103/PhysRevLett.92.083901} {\bibfield  {journal}
  {\bibinfo  {journal} {Phys. Rev. Lett.}\ }\textbf {\bibinfo {volume} {92}},\
  \bibinfo {pages} {083901} (\bibinfo {year} {2004})}\BibitemShut {NoStop}%
\bibitem [{\citenamefont {Beggs}\ \emph {et~al.}(2012)\citenamefont {Beggs},
  \citenamefont {Krauss}, \citenamefont {Kuipers},\ and\ \citenamefont
  {Kampfrath}}]{Beggs_Kampfrath_PRL108:033902_2012}%
  \BibitemOpen
  \bibfield  {author} {\bibinfo {author} {\bibfnamefont {D.~M.}\ \bibnamefont
  {Beggs}}, \bibinfo {author} {\bibfnamefont {T.~F.}\ \bibnamefont {Krauss}},
  \bibinfo {author} {\bibfnamefont {L.}~\bibnamefont {Kuipers}}, \ and\
  \bibinfo {author} {\bibfnamefont {T.}~\bibnamefont {Kampfrath}},\ }\href@noop
  {} {\bibfield  {journal} {\bibinfo  {journal} {Phys. Rev. Lett.}\ }\textbf
  {\bibinfo {volume} {108}},\ \bibinfo {pages} {033902} (\bibinfo {year}
  {2012})}\BibitemShut {NoStop}%
\bibitem [{\citenamefont {Kampfrath}\ \emph {et~al.}(2010)\citenamefont
  {Kampfrath}, \citenamefont {Beggs}, \citenamefont {White}, \citenamefont
  {Melloni}, \citenamefont {Krauss},\ and\ \citenamefont
  {Kuipers}}]{Kampfrath_Kuipers_PRA81:043837_2010}%
  \BibitemOpen
  \bibfield  {author} {\bibinfo {author} {\bibfnamefont {T.}~\bibnamefont
  {Kampfrath}}, \bibinfo {author} {\bibfnamefont {D.~M.}\ \bibnamefont
  {Beggs}}, \bibinfo {author} {\bibfnamefont {T.~P.}\ \bibnamefont {White}},
  \bibinfo {author} {\bibfnamefont {A.}~\bibnamefont {Melloni}}, \bibinfo
  {author} {\bibfnamefont {T.~F.}\ \bibnamefont {Krauss}}, \ and\ \bibinfo
  {author} {\bibfnamefont {L.}~\bibnamefont {Kuipers}},\ }\href {\doibase
  10.1103/PhysRevA.81.043837} {\bibfield  {journal} {\bibinfo  {journal} {Phys.
  Rev. A}\ }\textbf {\bibinfo {volume} {81}},\ \bibinfo {pages} {043837}
  (\bibinfo {year} {2010})}\BibitemShut {NoStop}%
\bibitem [{\citenamefont {Upham}\ \emph {et~al.}(2010)\citenamefont {Upham},
  \citenamefont {Tanaka}, \citenamefont {Asano},\ and\ \citenamefont
  {Noda}}]{Upham_Noda_APEX3:062001_2010}%
  \BibitemOpen
  \bibfield  {author} {\bibinfo {author} {\bibfnamefont {J.}~\bibnamefont
  {Upham}}, \bibinfo {author} {\bibfnamefont {Y.}~\bibnamefont {Tanaka}},
  \bibinfo {author} {\bibfnamefont {T.}~\bibnamefont {Asano}}, \ and\ \bibinfo
  {author} {\bibfnamefont {S.}~\bibnamefont {Noda}},\ }\href {\doibase
  10.1143/APEX.3.062001} {\bibfield  {journal} {\bibinfo  {journal} {Appl.
  Phys. Express}\ }\textbf {\bibinfo {volume} {3}},\ \bibinfo {pages} {062001}
  (\bibinfo {year} {2010})}\BibitemShut {NoStop}%
\bibitem [{\citenamefont {Nuccio}\ \emph {et~al.}(2010)\citenamefont {Nuccio},
  \citenamefont {Yilmaz}, \citenamefont {Wang}, \citenamefont {Wang},
  \citenamefont {Wu},\ and\ \citenamefont
  {Willner}}]{Nuccio_Willner_OL35:1819_2010}%
  \BibitemOpen
  \bibfield  {author} {\bibinfo {author} {\bibfnamefont {S.~R.}\ \bibnamefont
  {Nuccio}}, \bibinfo {author} {\bibfnamefont {O.~F.}\ \bibnamefont {Yilmaz}},
  \bibinfo {author} {\bibfnamefont {X.}~\bibnamefont {Wang}}, \bibinfo {author}
  {\bibfnamefont {J.}~\bibnamefont {Wang}}, \bibinfo {author} {\bibfnamefont
  {X.}~\bibnamefont {Wu}}, \ and\ \bibinfo {author} {\bibfnamefont {A.~E.}\
  \bibnamefont {Willner}},\ }\href {\doibase 10.1364/OL.35.001819} {\bibfield
  {journal} {\bibinfo  {journal} {Opt. Lett.}\ }\textbf {\bibinfo {volume}
  {35}},\ \bibinfo {pages} {1819} (\bibinfo {year} {2010})}\BibitemShut
  {NoStop}%
\bibitem [{\citenamefont {Sharping}\ \emph {et~al.}(2005)\citenamefont
  {Sharping}, \citenamefont {Okawachi}, \citenamefont {van Howe}, \citenamefont
  {Xu}, \citenamefont {Wang}, \citenamefont {Willner},\ and\ \citenamefont
  {Gaeta}}]{Sharping_Gaeta_OE13:7872_2005}%
  \BibitemOpen
  \bibfield  {author} {\bibinfo {author} {\bibfnamefont {J.}~\bibnamefont
  {Sharping}}, \bibinfo {author} {\bibfnamefont {Y.}~\bibnamefont {Okawachi}},
  \bibinfo {author} {\bibfnamefont {J.}~\bibnamefont {van Howe}}, \bibinfo
  {author} {\bibfnamefont {C.}~\bibnamefont {Xu}}, \bibinfo {author}
  {\bibfnamefont {Y.}~\bibnamefont {Wang}}, \bibinfo {author} {\bibfnamefont
  {A.}~\bibnamefont {Willner}}, \ and\ \bibinfo {author} {\bibfnamefont
  {A.}~\bibnamefont {Gaeta}},\ }\href {\doibase 10.1364/OPEX.13.007872}
  {\bibfield  {journal} {\bibinfo  {journal} {Opt. Express}\ }\textbf {\bibinfo
  {volume} {13}},\ \bibinfo {pages} {7872} (\bibinfo {year}
  {2005})}\BibitemShut {NoStop}%
\bibitem [{\citenamefont {Li}\ \emph {et~al.}(2008)\citenamefont {Li},
  \citenamefont {White}, \citenamefont {O'Faolain}, \citenamefont
  {Gomez-Iglesias},\ and\ \citenamefont {Krauss}}]{Li_Krauss_OE16:6227_2008}%
  \BibitemOpen
  \bibfield  {author} {\bibinfo {author} {\bibfnamefont {J.}~\bibnamefont
  {Li}}, \bibinfo {author} {\bibfnamefont {T.~P.}\ \bibnamefont {White}},
  \bibinfo {author} {\bibfnamefont {L.}~\bibnamefont {O'Faolain}}, \bibinfo
  {author} {\bibfnamefont {A.}~\bibnamefont {Gomez-Iglesias}}, \ and\ \bibinfo
  {author} {\bibfnamefont {T.~F.}\ \bibnamefont {Krauss}},\ }\href {\doibase
  10.1364/OE.16.006227} {\bibfield  {journal} {\bibinfo  {journal} {Opt.
  Express}\ }\textbf {\bibinfo {volume} {16}},\ \bibinfo {pages} {6227}
  (\bibinfo {year} {2008})}\BibitemShut {NoStop}%
\bibitem [{\citenamefont {Soref}\ and\ \citenamefont
  {Bennett}(1987)}]{Soref_Bennett_IEEEJQE23:123_1987}%
  \BibitemOpen
  \bibfield  {author} {\bibinfo {author} {\bibfnamefont {R.}~\bibnamefont
  {Soref}}\ and\ \bibinfo {author} {\bibfnamefont {B.}~\bibnamefont
  {Bennett}},\ }\href {\doibase 10.1109/JQE.1987.1073206} {\bibfield  {journal}
  {\bibinfo  {journal} {IEEE J. Quantum Electron.}\ }\textbf {\bibinfo {volume}
  {23}},\ \bibinfo {pages} {123 } (\bibinfo {year} {1987})}\BibitemShut
  {NoStop}%
\bibitem [{\citenamefont {Lepetit}\ \emph {et~al.}(1995)\citenamefont
  {Lepetit}, \citenamefont {Cheriaux},\ and\ \citenamefont
  {Joffre}}]{Lepetit_Joffre_JOSAB12:2467_1995}%
  \BibitemOpen
  \bibfield  {author} {\bibinfo {author} {\bibfnamefont {L.}~\bibnamefont
  {Lepetit}}, \bibinfo {author} {\bibfnamefont {G.}~\bibnamefont {Cheriaux}}, \
  and\ \bibinfo {author} {\bibfnamefont {M.}~\bibnamefont {Joffre}},\
  }\href@noop {} {\bibfield  {journal} {\bibinfo  {journal} {J. Opt. Soc. Am.
  B}\ }\textbf {\bibinfo {volume} {12}},\ \bibinfo {pages} {2467} (\bibinfo
  {year} {1995})}\BibitemShut {NoStop}%
\bibitem [{\citenamefont {Kampfrath}\ \emph {et~al.}(2009)\citenamefont
  {Kampfrath}, \citenamefont {Beggs}, \citenamefont {Krauss},\ and\
  \citenamefont {Kuipers}}]{Kampfrath_Kuipers_OL34:3418_2009}%
  \BibitemOpen
  \bibfield  {author} {\bibinfo {author} {\bibfnamefont {T.}~\bibnamefont
  {Kampfrath}}, \bibinfo {author} {\bibfnamefont {D.~M.}\ \bibnamefont
  {Beggs}}, \bibinfo {author} {\bibfnamefont {T.~F.}\ \bibnamefont {Krauss}}, \
  and\ \bibinfo {author} {\bibfnamefont {L.~K.}\ \bibnamefont {Kuipers}},\
  }\href {\doibase 10.1364/OL.34.003418} {\bibfield  {journal} {\bibinfo
  {journal} {Opt. Lett.}\ }\textbf {\bibinfo {volume} {34}},\ \bibinfo {pages}
  {3418} (\bibinfo {year} {2009})}\BibitemShut {NoStop}%
\bibitem [{\citenamefont {Vlasov}\ \emph {et~al.}(2005)\citenamefont {Vlasov},
  \citenamefont {O'Boyle}, \citenamefont {Hamann},\ and\ \citenamefont
  {McNab}}]{Vlasov_McNab_N438:65_2005}%
  \BibitemOpen
  \bibfield  {author} {\bibinfo {author} {\bibfnamefont {Y.~A.}\ \bibnamefont
  {Vlasov}}, \bibinfo {author} {\bibfnamefont {M.}~\bibnamefont {O'Boyle}},
  \bibinfo {author} {\bibfnamefont {H.~F.}\ \bibnamefont {Hamann}}, \ and\
  \bibinfo {author} {\bibfnamefont {S.~J.}\ \bibnamefont {McNab}},\ }\href
  {\doibase 10.1038/nature04210} {\bibfield  {journal} {\bibinfo  {journal}
  {Nature}\ }\textbf {\bibinfo {volume} {438}},\ \bibinfo {pages} {65}
  (\bibinfo {year} {2005})}\BibitemShut {NoStop}%
\bibitem [{\citenamefont {Melloni}\ \emph {et~al.}(2008)\citenamefont
  {Melloni}, \citenamefont {Morichetti}, \citenamefont {Ferrari},\ and\
  \citenamefont {Martinelli}}]{Melloni_Martinelli_OL33:2389_2008}%
  \BibitemOpen
  \bibfield  {author} {\bibinfo {author} {\bibfnamefont {A.}~\bibnamefont
  {Melloni}}, \bibinfo {author} {\bibfnamefont {F.}~\bibnamefont {Morichetti}},
  \bibinfo {author} {\bibfnamefont {C.}~\bibnamefont {Ferrari}}, \ and\
  \bibinfo {author} {\bibfnamefont {M.}~\bibnamefont {Martinelli}},\ }\href
  {\doibase 10.1364/OL.33.002389} {\bibfield  {journal} {\bibinfo  {journal}
  {Opt. Lett.}\ }\textbf {\bibinfo {volume} {33}},\ \bibinfo {pages} {2389}
  (\bibinfo {year} {2008})}\BibitemShut {NoStop}%
\bibitem [{\citenamefont {Adachi}\ \emph {et~al.}(2010)\citenamefont {Adachi},
  \citenamefont {Ishikura}, \citenamefont {Sasaki},\ and\ \citenamefont
  {Baba}}]{Adachi_Baba_IEEEJSTQE16:192_2010}%
  \BibitemOpen
  \bibfield  {author} {\bibinfo {author} {\bibfnamefont {J.}~\bibnamefont
  {Adachi}}, \bibinfo {author} {\bibfnamefont {N.}~\bibnamefont {Ishikura}},
  \bibinfo {author} {\bibfnamefont {H.}~\bibnamefont {Sasaki}}, \ and\ \bibinfo
  {author} {\bibfnamefont {T.}~\bibnamefont {Baba}},\ }\href {\doibase
  10.1109/JSTQE.2009.2032515} {\bibfield  {journal} {\bibinfo  {journal} {IEEE
  J. Sel. Topics Quantum Electron.}\ }\textbf {\bibinfo {volume} {16}},\
  \bibinfo {pages} {192 } (\bibinfo {year} {2010})}\BibitemShut {NoStop}%
\bibitem [{\citenamefont {O'Faolain}\ \emph {et~al.}(2010)\citenamefont
  {O'Faolain}, \citenamefont {Schulz}, \citenamefont {Beggs}, \citenamefont
  {White}, \citenamefont {Spasenovi\'{c}}, \citenamefont {Kuipers},
  \citenamefont {Morichetti}, \citenamefont {Melloni}, \citenamefont {Mazoyer},
  \citenamefont {Hugonin}, \citenamefont {Lalanne},\ and\ \citenamefont
  {Krauss}}]{OFaolain_Krauss_OE18:27627_2010}%
  \BibitemOpen
  \bibfield  {author} {\bibinfo {author} {\bibfnamefont {L.}~\bibnamefont
  {O'Faolain}}, \bibinfo {author} {\bibfnamefont {S.~A.}\ \bibnamefont
  {Schulz}}, \bibinfo {author} {\bibfnamefont {D.~M.}\ \bibnamefont {Beggs}},
  \bibinfo {author} {\bibfnamefont {T.~P.}\ \bibnamefont {White}}, \bibinfo
  {author} {\bibfnamefont {M.}~\bibnamefont {Spasenovi\'{c}}}, \bibinfo
  {author} {\bibfnamefont {L.}~\bibnamefont {Kuipers}}, \bibinfo {author}
  {\bibfnamefont {F.}~\bibnamefont {Morichetti}}, \bibinfo {author}
  {\bibfnamefont {A.}~\bibnamefont {Melloni}}, \bibinfo {author} {\bibfnamefont
  {S.}~\bibnamefont {Mazoyer}}, \bibinfo {author} {\bibfnamefont {J.~P.}\
  \bibnamefont {Hugonin}}, \bibinfo {author} {\bibfnamefont {P.}~\bibnamefont
  {Lalanne}}, \ and\ \bibinfo {author} {\bibfnamefont {T.~F.}\ \bibnamefont
  {Krauss}},\ }\href {\doibase 10.1364/OE.18.027627} {\bibfield  {journal}
  {\bibinfo  {journal} {Opt. Express}\ }\textbf {\bibinfo {volume} {18}},\
  \bibinfo {pages} {27627} (\bibinfo {year} {2010})}\BibitemShut {NoStop}%
\bibitem [{\citenamefont {Xu}\ \emph {et~al.}(2007)\citenamefont {Xu},
  \citenamefont {Dong},\ and\ \citenamefont {Lipson}}]{Xu_Lipson_NP3:406_2007}%
  \BibitemOpen
  \bibfield  {author} {\bibinfo {author} {\bibfnamefont {Q.}~\bibnamefont
  {Xu}}, \bibinfo {author} {\bibfnamefont {P.}~\bibnamefont {Dong}}, \ and\
  \bibinfo {author} {\bibfnamefont {M.}~\bibnamefont {Lipson}},\ }\href
  {\doibase 10.1038/nphys600} {\bibfield  {journal} {\bibinfo  {journal}
  {Nature Phys.}\ }\textbf {\bibinfo {volume} {3}},\ \bibinfo {pages} {406}
  (\bibinfo {year} {2007})}\BibitemShut {NoStop}%
\bibitem [{\citenamefont {Tanabe}\ \emph {et~al.}(2009)\citenamefont {Tanabe},
  \citenamefont {Notomi}, \citenamefont {Taniyama},\ and\ \citenamefont
  {Kuramochi}}]{Tanabe_Kuramochi_PRL102:043907_2009}%
  \BibitemOpen
  \bibfield  {author} {\bibinfo {author} {\bibfnamefont {T.}~\bibnamefont
  {Tanabe}}, \bibinfo {author} {\bibfnamefont {M.}~\bibnamefont {Notomi}},
  \bibinfo {author} {\bibfnamefont {H.}~\bibnamefont {Taniyama}}, \ and\
  \bibinfo {author} {\bibfnamefont {E.}~\bibnamefont {Kuramochi}},\ }\href
  {\doibase 10.1103/PhysRevLett.102.043907} {\bibfield  {journal} {\bibinfo
  {journal} {Phys. Rev. Lett.}\ }\textbf {\bibinfo {volume} {102}},\ \bibinfo
  {pages} {043907} (\bibinfo {year} {2009})}\BibitemShut {NoStop}%
\bibitem [{\citenamefont {Nguyen}\ \emph {et~al.}(2011)\citenamefont {Nguyen},
  \citenamefont {Sakai}, \citenamefont {Shinkawa}, \citenamefont {Ishikura},\
  and\ \citenamefont {Baba}}]{Nguyen_Baba_OE19:13000_2011}%
  \BibitemOpen
  \bibfield  {author} {\bibinfo {author} {\bibfnamefont {H.~C.}\ \bibnamefont
  {Nguyen}}, \bibinfo {author} {\bibfnamefont {Y.}~\bibnamefont {Sakai}},
  \bibinfo {author} {\bibfnamefont {M.}~\bibnamefont {Shinkawa}}, \bibinfo
  {author} {\bibfnamefont {N.}~\bibnamefont {Ishikura}}, \ and\ \bibinfo
  {author} {\bibfnamefont {T.}~\bibnamefont {Baba}},\ }\href {\doibase
  10.1364/OE.19.013000} {\bibfield  {journal} {\bibinfo  {journal} {Opt.
  Express}\ }\textbf {\bibinfo {volume} {19}},\ \bibinfo {pages} {13000}
  (\bibinfo {year} {2011})}\BibitemShut {NoStop}%
\bibitem [{\citenamefont {Tanabe}\ \emph {et~al.}(2010)\citenamefont {Tanabe},
  \citenamefont {Kuramochi}, \citenamefont {Taniyama},\ and\ \citenamefont
  {Notomi}}]{Tanabe_Notomi_OL35:3895_2010}%
  \BibitemOpen
  \bibfield  {author} {\bibinfo {author} {\bibfnamefont {T.}~\bibnamefont
  {Tanabe}}, \bibinfo {author} {\bibfnamefont {E.}~\bibnamefont {Kuramochi}},
  \bibinfo {author} {\bibfnamefont {H.}~\bibnamefont {Taniyama}}, \ and\
  \bibinfo {author} {\bibfnamefont {M.}~\bibnamefont {Notomi}},\ }\href
  {\doibase 10.1364/OL.35.003895} {\bibfield  {journal} {\bibinfo  {journal}
  {Opt. Lett.}\ }\textbf {\bibinfo {volume} {35}},\ \bibinfo {pages} {3895}
  (\bibinfo {year} {2010})}\BibitemShut {NoStop}%
\bibitem [{Note1()}]{Note1}%
  \BibitemOpen
  \bibinfo {note} {During the reviewing process, we became aware of
  arXiv:1110.5337v1, where an alternative method for achieving an indirect
  photonic transition \cite {Yu_Fan_NP3:91_2009} has been
  demonstrated.}\BibitemShut {Stop}%
\end{thebibliography}
%

\end{document}